\documentclass[12pt,preprint]{aastex}
\newcommand\beq{\begin{equation}}
\newcommand\eeq{\end{equation}}
\def\MSUN{\rm M$_{\odot}$}
\newcommand\Mdot {\dot{M}}

\begin{document}

\title{{\em Chandra} Observations of the Dwarf Nova WX Hyi in Quiescence} 

\author{Rosalba Perna\altaffilmark{1,2}, 
Jonathan McDowell\altaffilmark{2}, Kristen Menou\altaffilmark{3,4}   
John Raymond\altaffilmark{2}, Mikhail V. Medvedev\altaffilmark{5}}

\altaffiltext{1}{Harvard Society of Fellows, 74 Mount Auburn Street,
Cambridge, MA 02138, USA. e-mail: rperna@cfa.harvard.edu}

\altaffiltext{2}{Harvard-Smithsonian Center for Astrophysics, 60
Garden Street, Cambridge, MA 02138, USA; e-mail: jmcdowell@cfa.harvard.edu;
jraymond@cfa.harvard.edu}

\altaffiltext{3} {Celerity Foundation Fellow}

\altaffiltext{4} {Virginia Institute of Theoretical Astronomy,
Department of Astronomy, P.O. Box 3818, University of Virginia,
Charlottesville, VA, 22903, USA; e-mail: kmenou@virginia.edu}

\altaffiltext{5}{Department of Physics and Astronomy, Kansas University,
Lawrence, KS, 66045, USA; e-mail:medvedev@ku.edu}

\begin{abstract}
We report {\em Chandra} observations of the dwarf nova WX Hyi in
quiescence. The X-ray spectrum displays strong and narrow emission
lines of N, O, Mg, Ne, Si, S and Fe. The various ionization states
implied by the lines suggest that the emission is produced within a
flow spanning a wide temperature range, from $T\sim 10^6$ K to $T\ga
10^8$~K. Line diagnostics indicate that most of the radiation
originates from a very dense region, with $n\sim 10^{13}-10^{14}$~cm$^{-3}$.  
The {\em Chandra} data allow the first tests of
specific models proposed in the literature for the X-ray emission in
quiescent dwarf novae. We have computed the spectra for a set of
models ranging from hot boundary layers, to hot settling flows
solutions, to X-ray emitting coronae.  WX Hyi differs from other dwarf
novae observed at minimum in having much stronger low temperature
lines, which prove difficult to fit with existing models, and possibly a very
strong, broad O~VII line, perhaps produced in a wind moving at a few
$\times 10^3$ $\rm km~s^{-1}$.  The accretion rate inferred from the X-rays is
lower than the value inferred from the UV. The presence of
high-velocity mass ejection could account for this discrepancy while
at the same time explaining the presence of the broad O~VII line.  If
this interpretation is correct, it would provide the first detection of a
wind from a dwarf nova in quiescence.
\end{abstract}

\section{Introduction}

Cataclysmic Variables (CVs) are a class of interacting binaries in
which a donor star transfers mass onto a white dwarf (WD) accretor
(see Warner 1995 for a comprehensive review).  The characteristics of
the accretion flow depend mostly on the rate of accretion onto the WD
and on the WD magnetic field strength. In non-magnetic systems ($B\la
10^4$ G), accretion from the donor proceeds through an
undisturbed accretion disk, which connects to the WD through a
boundary layer (Patterson \& Raymond 1985).  When the WD is more
strongly magnetized, the accreting material is forced to follow the
topology of the field lines before it hits the surface (Aizu 1973).

As a dwarf nova of the SU UMa type, WX Hyi (discovered by Luyten 1932)
belongs to the class of non-magnetic CVs.  Its orbital period is
0.0748134 days, and because it does not show eclipses, only an
estimated inclination angle of $40\pm 10^\circ$ is available (Schoembs
\& Vogt 1981).  Schoembs \& Vogt (1981) derive a white dwarf mass of
0.9$\pm$0.3 \MSUN .  Normal outbursts occur on average every 11.2 days
(Ak, Ozkan \& Mattei 2002).  Distance estimates range from 120 pc
based on the correlation between H$\beta$ equivalent width and mass
transfer rate, $\Mdot$ (Patterson 1984), and 265 pc based on the
correlation between period and absolute magnitude during outburst
(Warner 1987).  We adopt 265 pc because the latter correlation is
tighter.  Patterson (1984) estimates $\Mdot = 1.8 \times 10^{16}~\rm
g~s^{-1}$ from the H$\beta$ equivalent width, which would give an
X-ray luminosity of $2 \times 10^{33}~\rm erg~s^{-1}$ if half the
accretion energy emerges as X-rays.  However, Schwope et al. (2002)
find an X-ray luminosity of only $4 \times 10^{31}~\rm erg~s^{-1}$
from the {\em ROSAT} All-Sky Survey data (assuming a 265 pc distance), which
yields $\Mdot\sim 3\times 10^{14}~\rm g~s^{-1}$.  K. Long (2003,
personal communication) finds accretion rates of order $10^{15}~\rm
g~s^{-1}$ from fits of accretion disk models to the UV continuum.  The
optical emission lines are single-peaked (Schoembs \& Vogt 1981),
perhaps placing WX Hyi in the SW Sex class of CVs.  Strong, rather
broad UV emission lines are seen in IUE spectra, with the N V line
being stronger than is typical of dwarf novae in quiescence (Hassall
et al.  1983).

Here we are interested in the structure of the accretion flow in the
quiescent state. In quiescence, the accretion rate in the disk is low
($\dot{M} \la 10^{16}$ g~s$^{-1}$) and the disk is too cool to contribute any
significant X-ray emission, while on the other hand the boundary layer
is expected to be optically-thin and thus emit copious amounts of
X-rays. According to the standard accretion theory (e.g. Lynden-Bell
\& Pringle 1974), the gravitational energy from the accreting material
should reemerge in roughly equal proportions from the disk and from
the boundary layer (BL). Whereas initial {\em Einstein} data on
quiescent CVs appeared to be consistent with emission from an hot,
optically-thin BL (Patterson \& Raymond 1985), later observations with
{\em Ginga, ROSAT, ASCA, RXTE} showed that the X-ray luminosity is
actually lower than theoretically expected.  Several suggestions
have been made to solve this ``mystery of missing BLs'' (e.g. Ferland
et al. 1982), such as (among others) reflection effects and cooling
flows (Done \& Osborne 1997), and disruption of the inner disk region
by the WD rotation (Ponman et al. 1995), by irradiation from the WD
(King 1997), or by ``coronal'' evaporation (Meyer \& Meyer-Hofmeister
1994).

The presence and structure of BLs (or more generally hot flows) in
quiescent CVs is therefore still a subject of debate, and wildly
different models have been proposed (e.g. Narayan \& Popham 1993,
Meyer \& Meyer-Hofmeister 1994; Mahasena \& Osaki 1999; Medvedev \&
Menou 2001).  While all the models can account for the {\em total}
observed X-ray luminosity, each of them, however, makes very specific
predictions for the physical conditions within the flow, and hence the
resulting temperature and density profiles.  As shown in a number of
papers (among the most recent ones: Pradhan 2000; Mauche et al. 2001;
Perna et al. 2000; Menou et al. 2001; Porquet et al. 2002; Dubau \&
Porquet 2002; Szkody et al. 2002), a powerful way to constrain the
characteristics of hot X-ray emitting flows is through the relative
intensities of emission lines produced within them.  The high
resolution of {\em Chandra} gives us the opportunity, for the first
time, to resolve relative line intensities and perform such an
analysis.

In this paper, we report {\em Chandra} observations of the
dwarf nova WX Hyi in quiescence.  Its X ray spectrum shows many lines
from different ionization states of N, O, Ne, Mg, Si, S and Fe. First,
we use line diagnostics to set overall constraints on the range of
temperatures and densities spanned by the flow. Second, we compute
X-ray spectra for different types of accretion flow structures, and
test them against our data.  The comparison shows that none of the
models provides an adequate fit to the entire emission line spectrum, in that
more plasma must be present at relatively low temperatures than
predicted by the models. A cooling flow model (used by Mukai et al. 2003
to model spectra of non-magnetic CVs) is consistent with the short-wavelength
spectra, but fails to reproduce the strength of the O~VII line.  
This line is much stronger and broader than in the objects studied
by Mukai et al., and we consider the intriguing possibility that
it might be produced in a wind from the accretion flow.

\section{Data Analysis}

WX Hyi was observed on-axis by the {\em Chandra} X-ray Observatory's
ACIS-S detector and HETG grating on 2002 Jul 25 and Jul 28
(observation ids 3721 and 2760) and with effective exposure times
after standard processing of 48366s and 49266s respectively.  Optical
data provided by the AAVSO show that WX Hyi was in quiescence during
both observations, with an outburst 4 days after the second
observation on August 2.  The previous detected outburst was 20 days
before our first observation.  Examination of the background light
curves shows no significant variability due to dropouts or flares. The
destreak algorithm was applied to remove electronic noise from the
noisier S4 chip, and the standard extraction parameters were used to
obtain dispersed count spectra. The $-1$ and $+1$ orders for both
observations were coadded to obtain summed HEG and MEG spectra.  There
was no significant difference in the spectra between the two
exposures. The secular degradation of the ACIS quantum efficiency was
taken into account using the acis-abs model while making the response
files.

The spectra were grouped to a signal-to-noise of 2 per bin.  Initial
attempts to model the lines in the merged spectrum using the CIAO
fitting program Sherpa using both global model fits and fits to
individual lines with a gaussian line model and polynomial fits to the
local continuum gave unsatisfactory, poorly constrained results for the
weaker lines. Instead, we determined line fluxes by simple integration
of the counts in a band around each line, subtracting the continuum
contribution using two methods: by comparison with counts in nearby
wavelength bands and by subtraction of a broken powerlaw fit to the
global continuum. These two methods gave consistent results, and are in
reasonable agreement with the gaussian line fits in the case of the
strongest lines.   In addition, measurements of the stronger lines in
the individual single order exposures give consistent results. Line flux
measurements are presented in Table 1 together with probable
identifications.  The errors in Table 1 are 1-sigma statistical uncertainties
computed for the line and continuum counts using the Gehrels
(1986) approximation suitable for low count Poisson statistics.

The cleanest X-ray line is the O~VIII resonance line at 18.967 \AA .
Its width is  $\sim 900~\rm km~s^{-1}$ (FWHM) after correction for the
instrument profile.  Correcting for the $40^\circ$ inclination, the
intrinsic width is $\sim 1400~\rm km~s^{-1}$.  While this is similar to the widths 
($\sim 1200~\rm km~s^{-1}$) of
the UV lines, the wings ($\sim 4000~\rm km~s^{-1}$)
apparent in the UV line profiles (K. Long
2003, private communication) are absent.  The inclination-corrected
line width is $\la 1/3$ the Keplerian velocity near the WD surface.

\section{Constraints on the  characteristics of the accretion flow 
from the X-ray spectrum}

\subsection{Line diagnostics}

The X-ray spectrum, including both the MEG and HEG observations, is
displayed in Figure 1.  The spectrum shows a smooth continuum with
strong emission lines, particularly from the H and He-like ions of O,
Mg, Ne, Si, Fe, as well as from all the Fe L-shell complex (Fe~XVII-Fe~XXIV).  
The presence of this variety of ionization states indicates
emission from a plasma distributed over a wide range of temperatures,
from $T\sim 10^6$ K, necessary to have some contribution to the O~VII
line, up to $T\sim 10^8$ K, a temperature at which the ion Fe~XXVI is
most abundant in equilibrium (without strong photoionizing
radiation).

From the X-ray spectrum in Figure 1, one can see that the O~VII line
profile appears to be wider that those of O~VIII and the other lines,
though it is clearly noisy.  The profile, which is separately
displayed in Figure 2, seems to consist of narrow O~VII resonance and
forbidden lines on top of a broad feature perhaps 5000 $\rm km~s^{-1}$
wide.  Some of the noise level results from the presence of bad
columns in the $+1$ order of the MEG, which are broadened by
dithering.  However, careful examination of the counts in the $-1$
order in the individual exposures indicates that the broad feature is
probably real, though details of the structure are questionable.  Note
that the effective area changes rapidly in this wavelength range,
making the number of counts in the blue-shifted part of the feature
more significant. It is difficult to determine an unambiguous value
for the uncertainty, in that different choices for spectral binning
yield different estimates of the significance of the feature.  There
are several lines of calcium near this wavelength, but comparison with
iron lines formed at the same temperatures indicates that the calcium
lines are too weak to account for the observed structure. It should be
noted, however, that there appears to be also a feature around 20.3~\AA 
\ for which we have no identification.
  
If the broad O VII component is genuine, 
the lack of broad components in the other lines
suggests that the high velocity material only appears in low
ionization species.  The lack of this component in N~VII must be due
to the small number of counts in that line.  The line width is
comparable to or larger than the widths ($\sim 1200~\rm km~s^{-1}$)
of the UV Lines (K. Long 2003,
personal communication), so the broad O~VII might originate in the
innermost part of the disk or in a wind.

As a first attempt to constrain the physical conditions in the
accretion flow, we use the diagnostics provided by the He-like triplet
line ratios, which have been widely used in analyses of the solar
plasma (Gabriel \& Jordan 1969; Mewe \& Schrijver 1978; Doyle 1980;
Pradhan \& Shull 1981).  In particular, the ratio $R\equiv f/i$
between the forbidden line $f$ and the intercombination line $i$ is a
strong function of the density at high densities, due to the
suppression of the forbidden lines with increasing density. The ratio
$R$ is typically on the order of a few at low densities, and rapidly
drops at densities larger than a critical value $n_c$ that increases
with $Z$, ranging from $n_c\approx 6\times 10^8$~cm$^{-3}$ for C, to
$n_c\approx 3\times 10^{17}$~cm$^{-3}$ for Fe (e.g. Porquet et
al. 2001).  We find the 1$\sigma$ upper limits $R<0.47$ and $R<0.92$ for
Ne\footnote{It should be noted that the intercombination line of Ne is
not very clean, and might have some contamination from Fe XIX which
is not possible to quantify.}
and Mg respectively.  The 1s2s$^3S$ - 1s2p$^3P_1$ transition,
however, is driven not only by collisions, but also by UV radiation,
so that in gas near a 30,000 K black body the $f/i$ ratio will be in
the high density limit for elements through Mg (Mauche 2002).  We do
not know the white dwarf temperature for WX Hyi.  K. Long (private
communication, 2003) has fit UV continua of WX Hyi observed by the
{\em Hopkins Ultraviolet Telescope} and by the {\em Hubble Space
Telescope} GHRS with combinations of white dwarf and accretion disk
spectra, and typically finds WD temperatures above 20,000 K.  The
$f/i$ ratios of O~VII, Ne~IX and Mg~XI are probably dominated by UV
radiation, but the $f/i$ ratio for Si~XIII should be dominated by
collisions, even if the entire UV continuum in the Hopkins Ultraviolet
Telescope wavelength range arises from the white dwarf.
Unfortunately, we are unable to obtain a reliable measurement of the
Si~XIII ratio.

Another useful plasma diagnostic is the ratio $G\equiv (i+f)/r$, where
$r$ is the resonance line. This is sensitive to the ionization state
of the gas and to the electron temperature.  When the resonance line
is strong compared to the forbidden and the intercombination line
($G\la 1$), the plasma is collision-dominated. On the other hand, a
plasma in which photoionization is important will have a weaker
resonance line. Values of $G>4$ are typically considered to be
indicative of a photoionization-dominated plasma (e.g. Porquet et
al. 2001).  For O, Si and S, we found that the $i$, $r$ and $f$ lines
were not well resolved to allow a reliable estimate of the $G$
value. In the cases of Mg and Ne, we were able to set the 1$\sigma$
upper limits $G<0.73$ and $G<1.08$, respectively. This indicates that
photoionization is not dominant.

A useful density diagnostic is provided by the ratio between the 17.10
\AA \ and the 17.05 \AA \ lines of the ion Fe~XVII, which is less
sensitive to UV radiation than are the He-like line ratios (Mauche et
al. 2001). The Fe~XVII ratio $I(17.10)/I(17.05)$ is found to be $\la
0.6$ (at the $1\sigma$ level) for WX Hyi. For a
collisionally-dominated plasma, this value suggests a density $n\ga
3\times~10^{13}\rm ~cm^{-3}$ in the relatively cooler region ($T\sim
3$--$6\times 10^6$ K) where Fe~XVII is produced. A comparable density
is implied by the best value of the ratio between the 11.92 \AA \ and
the 11.77 \AA \ lines of Fe XXII, which is found to be $0.68\pm 0.4$ in
our data.  Accounting for the error in the measurement, the $1\sigma$
confidence level for the density ranges between $10^{13}-10^{14}~\rm
cm^{-3}$.  As discussed by Mauche, Liedahl \& Fournier (2003), this
density determination is very insensitive to temperature and
photoexcitation.

\subsection{Comparison with specific models for the X-ray emission}

The physics, and hence the density-temperature structure, of boundary
layers is still poorly understood. Only a few models have been
proposed, and they have been primarily tested on their ability to account for the
{\em total} observed X-ray and UV radiation. The {\em Chandra}
observations of the X-ray emission, by resolving lines,
allow us to set much tighter constraints on the models, due to the
high sensitivity of the relative line intensities to the temperature
and density profile characterizing the flow.  Here we consider five
types of models proposed in the literature to explain the origin of
the quiescent X-ray emission of CVs. For each of them, we compute the
expected X-ray spectrum using the temperature and density profiles
that they predict, and compare the spectra with our observations.
The cooling flow model has been implemented in the
{\bf xspec} package (Arnaud 1996). For the other models we compute the
emissivity of the medium by using the atomic rate packages of the
shock model code described by Raymond (1979), modified with updated
atomic rates, as in Cox \& Raymond (1985). This package has the
disadvantage that some Fe L-shell multiplets that are resolved in our
spectra are computed as single lines.  The package also lacks recent
improvements to the atomic rates of the more complex Fe ions.  It
computes an ionization balance which is quite similar to that of
Mazzotta et al. (1998) in ionization equilibrium.  The advantages of
the code are that it allows us to compute time-dependent ionization
and to include photoionization.  It includes dielectronic
recombination satellite lines and the contributions of recombination
to excited levels for the spectra of H-like and He-like ions.
Overall, the predictions should be reasonably accurate for most of the
strong lines; those of H-like and He-like ions, Fe~XVII and Fe~XXIV.
For the models presented here we included the effects of
photoionization by the X-ray radiation produced within the flow
itself, though this turned out to have little effect on the emission
lines.  For a proper comparison with the data, the computed X-ray
spectra were then convolved with the response matrices and effective
areas of the {\em Chandra} detectors. In particular, we used the HEG
detector for $\lambda\le3$ \AA, and the MEG detector for $\lambda >
3$ \AA.

(a) {\em Cooling flow}

A cooling flow model, based on Mushotzky \& Szymkowiak (1988), has
been implemented in {\bf xspec} with the routine {\bf mkcflow}.  Its
basic assumption is of steady-state isobaric radiative cooling. The
two main parameters are the maximum temperature, $T_{\rm max}$, and
the overall normalization, which directly relates to the mass
accretion rate.  Mukai et al (2003) showed that cooling flow models
provide a good representation of the X-ray spectral properties of
non-magnetic CV systems.  A fit to our data with the model {\bf
mkcflow} is shown in the top panel of Figure 3. We find that the fit
is acceptable at short wavelengths but poorly constrained, with the
maximum temperature parameter at a value of 20 keV, uncertain by a
factor of two, and the column density $N_{\rm H}=(2\pm 2) \times
10^{20}$ cm$^{-2}$ (basically consistent with zero).  The value of the
temperature is consistent with the Fe XXVI/Fe XXV ratio that gives an
upper limit of 10 keV. Note that this ratio is about 1/7 in WX Hyi as
opposed to 3/4 in U Gem (Szkody et al. 2002), implying a corresponding
lower temperature of WX Hyi.  The value of $\dot{M}$ implied by the
cooling model is $\sim 2\times 10^{14} \rm g~s^{-1}$  for $T_{\rm max}=20$ keV
(assuming that half of the X-rays are absorbed by the WD).  This is
about a factor of 5 smaller than the value found by K. Long (2003,
private communication) from fits of disk models to the UV continuum.
While the system is by definition variable, it is not surprising that
the X-ray luminosity is less than expected from $\dot{M}$ farther out
in the disk.  Possible interpretations are that $\dot{M}$ decreases
inward in the disk as material builds up for the next outburst, that
some material and energy is lost to a wind, or that some energy is
advected to the WD surface and reemitted in the UV.

Overall, the cooling flow model appears to account reasonably well for the
continuum and the line strengths in the short wavelength region of the
spectrum, but it underpredicts the emission at long wavelengths. In
particular, it is not able to account for the emission from the O~VII
line, which in WX Hyi is significantly stronger than in the objects studied by
Mukai et al. (2003).

(b) {\em Hot boundary layer}

Narayan \& Popham (1993; NP93) describe detailed models for thin disk
boundary layers in CVs. A key element of their theory is the role of
heat advection which, they show, allows boundary layers at low
accretion rates to be radially thicker and significantly hotter than
those at higher accretion rates.  We have used the profiles of density
and temperature corresponding to the model with $\dot M=2 \times
10^{15} \rm g~s^{-1}$ , which is the closest to the accretion rate in our
system.  The behavior of the low temperature region closest to the
white dwarf surface cannot be read from the plots in NP93, but this is
a small fraction of the total X-ray luminosity and may be neglected.  We
note that the comparison with this particular solution should be
interpreted with caution, as temperature and density do not simply
scale with accretion rate.  Panel (b) of Figure 3 shows a comparison
between our data and the X-ray spectrum that we computed for this
boundary layer solution.  The spectrum has been rescaled to match the
continuum. If the density profile were to scale linearly with the
accretion rate, the data would imply a value of $\dot{M}$ of about
$2\times 10^{14}$ g~s$^{-1}$.  As the figure shows, this model predicts substantially
more emission than the cooling flow in the O~VII and O~VIII lines.  This
is a result of the fact the BL solution of NP93 predicts an increase
of the density in the outer, colder regions of the flow where the
oxygen emission peaks. However, the reasonable agreement between the
observed and predicted low temperature lines is partly fortuitous in
that Narayan \& Popham assumed bremsstrahlung cooling, while the
actual line cooling rate at the temperatures where the O~VII, O~VIII and 
Fe~XVII lines are formed is nearly an order of magnitude higher.  Thus if
Narayan \& Popham had used a cooling rate consistent with the lines we
observe, their model would have predicted weaker emission by 
an order of magnitude.

(c) {\em Coronal siphon flow}

Meyer \& Meyer-Hofmeister (1994) investigated a model based on the
interaction between a cool disk and a hot corona above it via thermal
conduction. They showed that, in these conditions, a coronal ``siphon
flow'' can develop. This evaporates mass from the disk, which is
partly accreted onto the white dwarf and partly lost to a wind. The
solutions for this model are uniquely determined once the viscosity
parameter $\alpha$ for the flow is specified\footnote{Under the
conditions in which the radiative energy loss in the
corona is small compared to conductive energy loss towards
the lower boundary and the wind loss, the radial solutions
for $T$ and $\rho$ are independent of $\dot{M}$ for a given
white dwarf mass (Meyer \& Meyer-Hofmeister 1994).}.  We varied the value of
this parameter to find the best fit coronal flow solution to our data.
The resulting X-ray spectrum predicted for the coronal siphon flow is
shown in the middle panel of Figure 3.  As it can be seen, this model
falls short of producing enough emission in the long wavelength region
of the spectrum. This is a consequence of the fact that this solution
predicts a radial temperature profile $T\propto r^{-1}$, and a run of
density which is a very steep function of the radius, $n\propto
r^{-3}$.  This implies that there is relatively little mass at the
lower temperatures where the longer wavelength emission is generated.

(d) {\em X-ray emitting corona}

Mahasena \& Osaki (1999) constructed steady-state solutions of X-ray
emitting corona, including the effect of thermal conduction.  We have
used the density and temperature profiles shown in their Figure~1 for
an accretion rate $\dot M =10^{15}$ g~s$^{-1}$, and calculated the
corresponding X-ray spectral emission.  It was not possible to read
numbers from their plot for the region below $6 \times 10^6$ K, so we
computed a matching solution that neglected the rotation and infall
velocities, both of which are small in this region.  The result is
compared to the data in panel (d) of Figure 3.  This comparison should
also be taken with caution because the value of the accretion rate is
not precisely tuned to match that of WX Hyi.  This coronal model,
which transports most of the accretion luminosity to a cool, thin,
conduction dominated layer near the white dwarf surface and radiates
it there, produces emission lines that are far too strong and a
continuum that is much too soft.  The higher accretion rate model of
Mahasena \& Osaki, $\dot{M} =10^{16}$ g~s$^{-1}$, did a much better job
matching the {\em Chandra} spectrum of U Gem (Szkody et
al. 2002)\footnote{Note that there is a typo in the Szkody et
al. (2002) in reporting the higher value of the accretion rate used as
$\dot{M} =10^{15}$ g~s$^{-1}$ rather than $\dot{M} =10^{16}$ g~s$^{-1}$.}  because
it is closer to the cooling flow solution, which matches that spectrum
quite well (Mukai et al. 2003).  However, the cooling model predicts
too little emission in the low temperature lines to match the WX Hyi
spectrum.  The density in the region where the Fe XXII lines are
produced is about $3 \times 10^{14}~\rm cm^{-3}$ in this model, a few
times higher than the density sensitive line ratio indicates.

(e) {\em Hot settling flow solutions} 

Following Medvedev \& Narayan (2001), Medvedev \& Menou (2002)
presented solutions for hot accretion onto unmagnetized, rotating
white dwarfs.  Together with the accretion rate, the WD rotation rate
is an important parameter in these models, because viscously-mediated
losses of rotational energy by the WD constitute an additional source
of energy powering the X-ray emission from the flow (see Medvedev \&
Menou 2002 for examples). To calculate the hot flow models we used the
numerical relaxation code which solves height-integrated
two-temperature hydrodynamic equations with high spatial resolution
providing accurate solutions even deep inside the boundary layer. This
is the same code which has previously been used by Medvedev \& Menou
(2002) except for the cooling part which now also incorporates (in
addition to bremsstrahlung) the emission line cooling in the form
$q_{\rm line}^-=6.6\times10^{-22} T_5^{-0.73} n_e
n_H$~erg~cm$^{-3}$~s$^{-1}$, which is a good approximation above
$T=10^5$~K, where $T_5=T/(10^5~{\rm K})$ (Raymond, Cox \& Smith 1976).  
At the high densities of the WX Hyi boundary layer, quenching of UV lines
reduces the cooling rate below $10^6$ K, but here we are concerned only with
lines formed above $10^6$ K. 
In our present solutions we
have chosen the outer boundary of the flow to be at radius $R=10^2
R_{\rm WD}$, which is realistic.  We have computed hot settling flow
models appropriate for WX Hyi. Absent any observational constraint on
the WD rotation in this system, we have computed a variety of models
with values for the spin parameter, $s$ (the angular velocity of the
WD in units of the Keplerian angular velocity), ranging from 0.03 to
0.3. The accretion rate in the models was adjusted so that the
observed X-ray luminosity is reproduced in each case (see Medvedev \&
Menou 2002 for details). The required value was on the order of
$10^{15}$ g~s$^{-1}$.

We found that the X-ray spectrum predicted by the hot settling flow
solution greatly overpredicts the X-ray line emission with respect to
our observations. Figure 3 (e) shows the spectrum predicted by this
solution for the case $s=0.03$. As $s$ increases, line emission
becomes even more pronounced. Therefore, unless optical depth effects
become sufficiently important to suppress line emission, this solution does
not appear to reproduce the observations well.

\section{Conclusions}

The {\em Chandra} observations of the dwarf nova WX Hyi in quiescence
show resolved spectral lines produced within the accretion flow.  Line
diagnostics have allowed us to set constraints on the characteristics
of the hot flow, showing that gas densities are very high,
$n\sim 10^{13}$--$10^{14}~\rm cm^{-3}$, and span a wide range of
temperatures, $T\sim 10^6$--$10^8$ K.

The spectrum of WX Hyi shows some unusual features if compared to the
spectra of the objects in the same class studied by Mukai et
al. (2003).  Lines in the longer wavelength region of the spectrum are
relatively stronger. The O~VII line especially  cannot be
accounted for by the cooling flow model used by Mukai et al. to fit
the spectra of this type of objects.

The {\em Chandra} data allow the first tests of specific models
proposed in the literature for the X-ray emission in quiescent dwarf novae. We
have computed the spectra for a set of models ranging from hot
boundary layers, to hot settling flows solutions, to X-ray emitting
coronae. While most of these models can reproduce well the shape of
the continuum, none of them can fully account for the relative line
strengths over the entire spectral range.  Most coronal models fall
short of predicting enough radiation at longer wavelengths, though a
thermal conduction-dominated boundary layer predicts too much long
wavelength emission.  Hot accretion solutions tend to overpredict the
short wavelength emission.  It is possible, but far from obvious, that
some combination of these models, perhaps with thermal conduction
present but reduced by a magnetic field, might match the observations.

Given the difficulties of the existing models, it is possible that
somewhat different physical processes dominate the boundary layer
flow.  In particular, it has been suggested that magnetic flux tubes
rise above the disk surface and produce X-rays in the same manner as
solar flares (Galeev, Rosner \& Vaiana 1979).  While X-ray emission
from the disk as a whole is excluded by the X-ray light curve of OY
Car, which limits much of the X-ray emission to the immediate vicinity
of the white dwarf (Ramsay et al. 2001), the boundary layer is likely
to be magnetically dominated.  The boundary layer itself is stable
against the magnetorotational instability (Balbus \& Hawley 1991), but magnetic flux generated
just beyond the boundary layer diffuses into it, and the toroidal
component is strongly amplified by the enormous shear (Armitage 2002;
Steinacker \& Papaloizou 2002).  Lacking a more definite prediction,
one might expect a power law distribution of X-ray flare energies, as
on the Sun (UeNo et al. 1997) and perhaps a significant contribution
from flares that reach modest temperatures (e.g. Raymond 1990).
Detailed models require Monte Carlo simulations of flare energies,
intervals between flares, and physical processes such as thermal conduction,
and are beyond the scope of this paper. It is possible, however, that
some combination of parameters can be found to roughly match the data.

It is therefore possible that the boundary layer is better described
as a collection of impulsive magnetic reconnection events than as a
smooth fluid flow.  An implication of this idea is that a significant
amount of mass may be expelled as a wind.  Solar flares are often
accompanied by Coronal Mass Ejections (CMEs), and on average the CMEs
carry somewhat more kinetic energy than the flares emit as radiation.
Thus mass ejections could easily account for the factor of 3 or so
discrepancy between the accretion rates estimated from the UV spectrum
and the X-ray spectrum.  Thus far, winds from cataclysmic variables
have only been detected in high $\dot{M}$ systems.  It has recently
been shown that these winds are not driven by radiation pressure alone
(Mauche \& Raymond 2000; Proga 2003). There is as yet no way to
determine whether large scale fields, as in Proga's models, or small
scale CME-like eruptions dominate the wind.  Winds from low accretion
rate systems are difficult to detect because of the high ionization
state expected.

Assuming the broad component of the O~VII emission (section 2) is
real, it could be formed in the wind hypothesized in the previous
paragraph.  A simple estimate of the mass loss rate from a surface
corresponding to the white dwarf circumference and the boundary layer
width ($\sim 10^8$ cm) with a density $n\sim 3 \times 10^{13}~\rm
cm^{-3}$ (as required to match the O~VII luminosity) and a speed $v$
of 3000 $\rm km~s^{-1}$ yields about $2 \times 10^{15}~\rm g~s^{-1}$,
or more than the total $\Mdot$ inferred from the UV spectra.  Thus in
order to identify the broad O~VII emission with a wind, we require a
filling factor of order 0.1. With this filling factor one would in
fact need a density $\sim\sqrt{10}$ larger to keep the emission
measure ($\propto n^2\, V$) of the O~VII line at the observed
value. The accretion rate $\dot{M}\propto n\,v\,A$ ($A$ being the
area) would then be $\sim\sqrt{10}$ smaller, consistent with the value
inferred from the UV observations.  One must be cautious about a wind
interpretation both because of the low statistical quality of the
O~VII profile and because a wide O~VII profile might arise from
Keplerian rotation of the inner disk just outside the boundary layer.
However, a wind of moderate ionization state does provide an appealing
explanation for the difference between the accretion rates derived
from UV and X-ray observations.

\bigskip

We thank the referee, Chris Mauche, for his careful review of our
manuscript and very useful comments.  In this research, we have used,
and acknowledge with thanks, data from the AAVSO International
Database, based on observations submitted to the AAVSO by variable
star observers worldwide.  This analysis was supported by the {\em
CHANDRA} grant GO2-3032X to the Smithsonian Astrophysical Observatory
and by the Celerity Foundation.

\clearpage

\begin{deluxetable}{lccc}
\tablewidth{0pt}
\tablecaption{Line summary}
\tablehead{\colhead{Element} & \colhead{$\lambda$ \tablenotemark{(a)}} & 
\colhead{HEG flux\tablenotemark{(b)}} 
& \colhead{MEG flux\tablenotemark{(b)}}}
\startdata		    
  Fe XXVI	&   1.78 &    $12.5\pm 6.4$ &   $-$ \\
  Fe XXV	&  1.85 &     $54.1\pm 7.1$ & $67\pm 20$ \\
  Fe K$\alpha$       &  1.94 &      $10.0\pm 3.9$  & $-$ \\
 S XVI &  4.73 &	      $4.5\pm 1.8$ &  $4.1\pm 0.9$ \\
 S XV &      5.05 &   $< 6.0$   & $ 2.8\pm 1.1$ \\  
 Si XIV & 6.19 &	$6.3\pm 0.9$ &  $5.6\pm 0.5$ \\
 Si XIII	& 6.65 &    	 $1.0\pm 0.7$ &  $0.73\pm 0.44$\\
 Fe XXIV &  8.00	&       $1.4\pm 0.4$ &  $0.5\pm 0.2$\\
 Mg XII & 8.42 &         $2.1\pm 0.4$ &  $2.3\pm 0.3$ \\
Mg XI & 9.17 &                $<$1.0 & $0.63\pm 0.3$\\
Mg XI & 9.23 &                $-$ & $0.24\pm 0.1$\\
Mg XI & 9.31 &               $-$ & $0.22\pm 0.2$\\
Ne X & 10.24 &             $1.4\pm 1.5$ & $0.5\pm 0.2$\\
Fe XXIV& 10.62\tablenotemark{(c)}  &    $1.9\pm 0.6$ & $1.8\pm 0.3$\\
Fe XXIV &10.66\tablenotemark{(c)}  &   $0.9\pm 0.4$ & $1.7\pm 0.3$\\
Fe XXIV & 11.05      &   $1.3\pm 0.7$ & $0.85\pm 0.3$\\
Fe XXIV &11.17        & $2.4\pm 0.8$ & $2.3\pm 0.3$\\
Fe XXIII & 11.73 & $1.5\pm 0.7$   & $1.1\pm 0.2$\ \\
Fe XXII & 11.77 & $1.3\pm 0.7$   & $1.1\pm 0.3$\\
Fe XXII &11.92     &    $<$ 1.3   &  $0.75\pm 0.4$\\
Ne X &12.13          &   $4.3\pm 1.0$  & $5.0\pm 0.6$\\
Ne IX &13.44          &   $-$  & $2.3\pm 0.5$\\
Ne IX &13.55          &   $-$  & $1.7\pm 0.5$\\
Ne IX &13.70          &   $-$  & $<0.8$\\
Fe XVII &15.03   &       $-$  &  $2.6\pm 0.6$\\
Fe XVII  &17.05    &    $-$  &  $1.9\pm 0.6$\\
Fe XVII & 17.10      &  $-$  & $<$ 1.1\\
O VIII &18.96       &   $-$   & $7.0\pm 1.3$  \\
O VII &21.60\tablenotemark{(d)}      &   $-$   & $3.9\pm 2.1$ \\
O VII & 21.80\tablenotemark{(d)}    & $-$ & \\
Fe XXIV& 21.97\tablenotemark{(d)} & $-$ & \\
N VII & 24.74    &       $-$ &    $4.5\pm 1.5$\\
\enddata
\tablenotetext{a}{Units are \AA\,. The center of the lines
are taken at the theoretical values obtained from the ATOMDB database
distributed with {\em CIAO}, except for the blend of Fe XXII and Fe XXIII near 11.7 \AA $\,$
where we use wavelengths measured from solar spectra (Doschek \& Cowan 1989).}
\tablenotetext{b}{Units are $10^{-14}$ erg/cm$^2$/s\,.} 
\tablenotetext{c}{These two lines are blended in MEG\,.}
\tablenotetext{d}{These three lines are blended\,.}
\end{deluxetable}

\clearpage

\begin{figure}[t]
\plotone{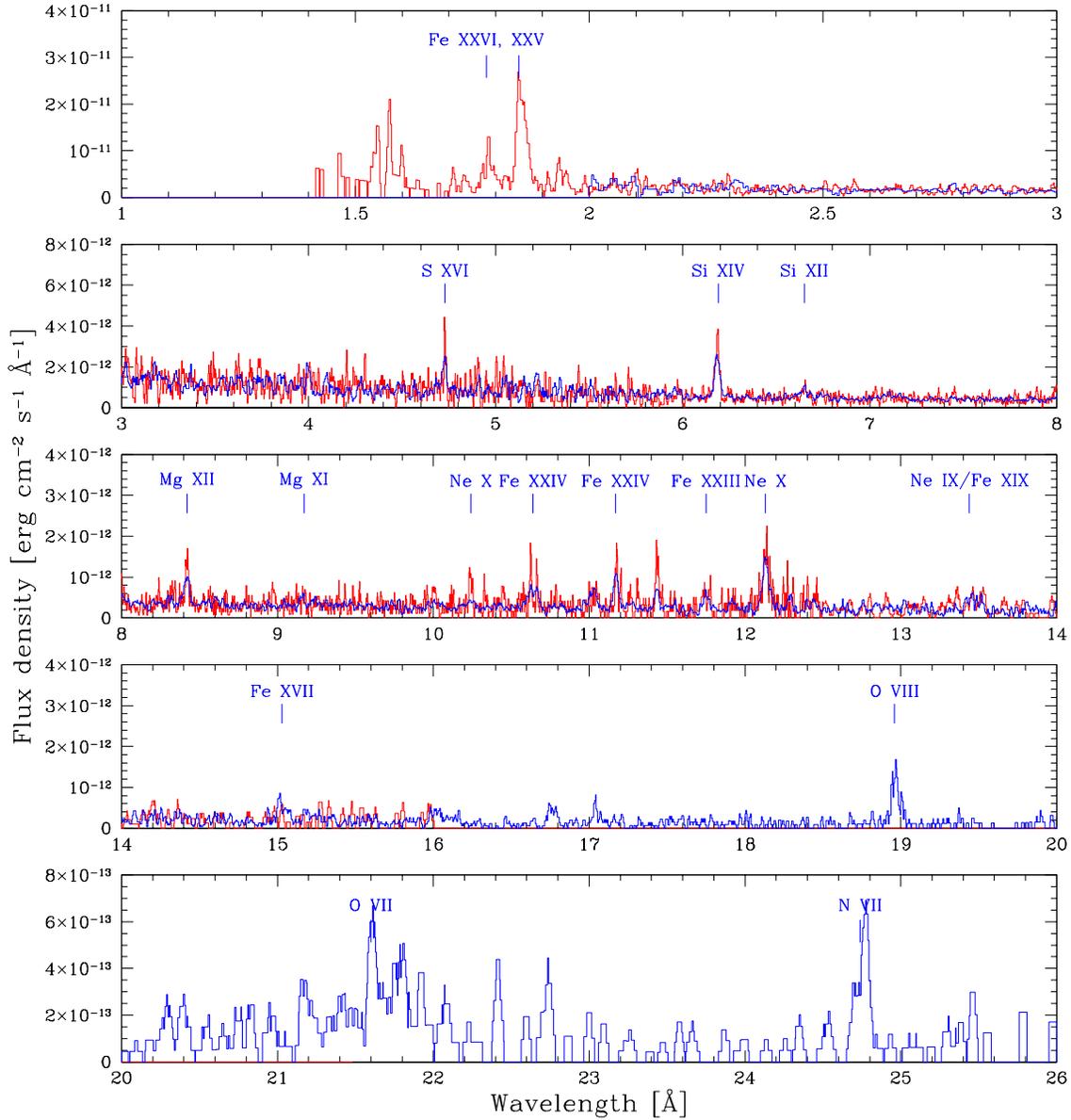}
\caption{{\em Chandra} spectrum of the dwarf nova WX Hyi in
quiescence. The spectrum has been binned so that each bin has a
S/N=2. Both the MEG (blue) and the HEG (red) spectra are displayed.}
\end{figure}

\begin{figure}[t]
\plotone{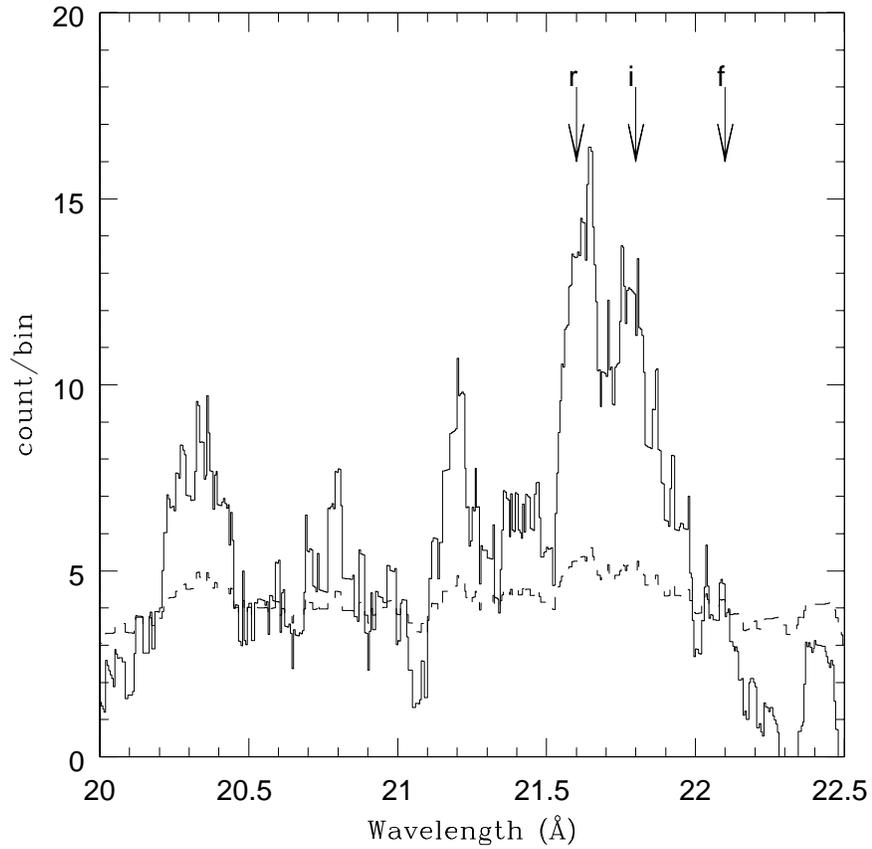}
\caption{Details around the O VII line complex, showing flux smoothed with 
a 0.125A boxcar (solid line). 
The dashed line indicates combined uncertainty     
from background and count statistics. The r, i and f
lines of O VII are indicated. Excess flux between 21.0 and
21.5 \AA \ appears real and may represent blueshifted
velocity components of O VII.}
\end{figure}

\begin{figure}[t]
\plotone{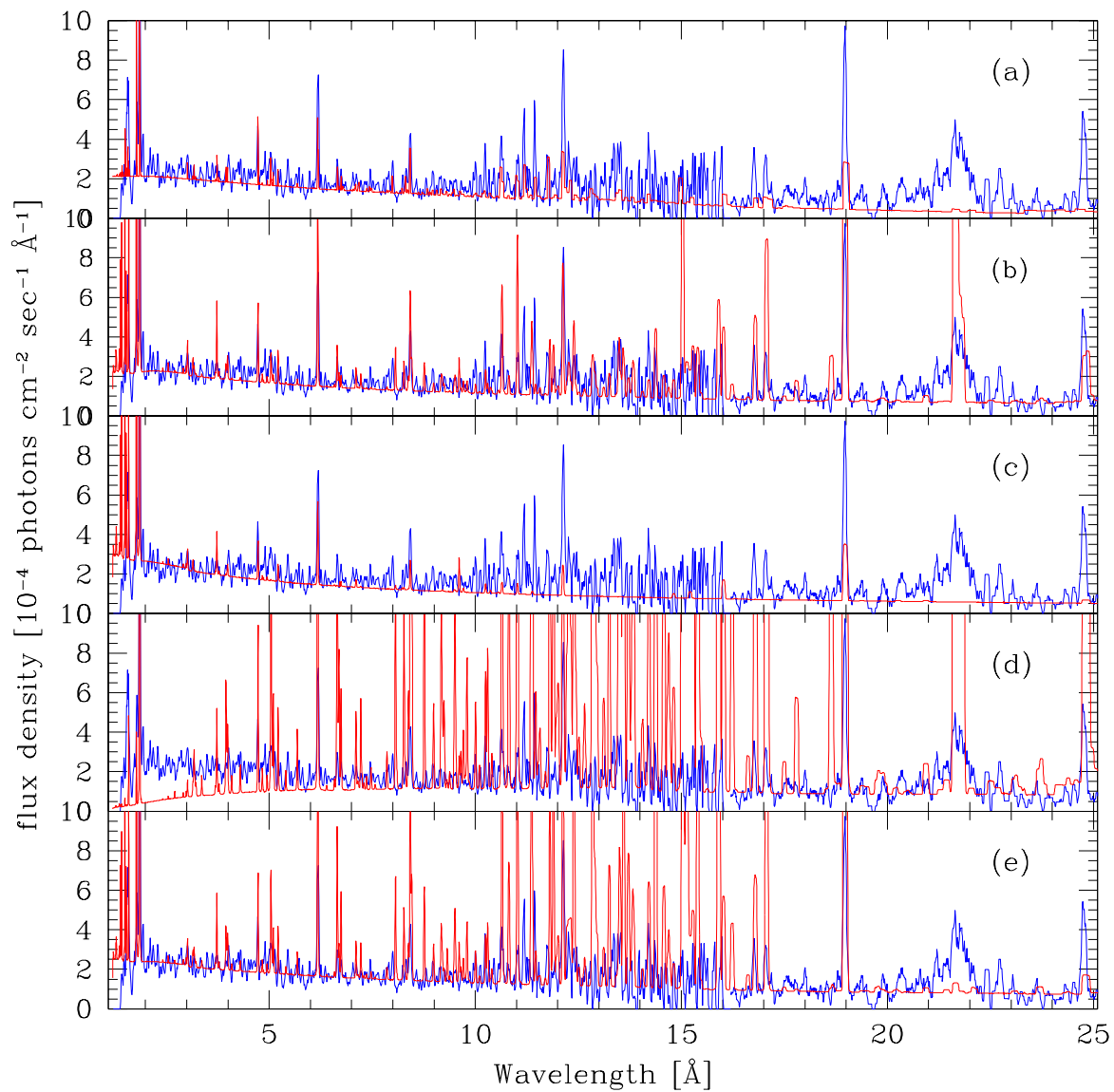}
\caption{The quiescent emission of WX Hyi as seen by {\em Chandra}
(blue), compared to theoretical spectra computed for several
models (red).  From top to bottom: panel (a): cooling flow
(Mushotzky \& Szymkowiak 1988); (b): hot boundary layer (Narayan \&
Popham 1993); (c): coronal siphon flow (Meyer \& Meyer-Hofmeister
1994); (d) hot X-ray emitting corona (Mahasena \& Osaki 1999); (e) hot
settling flows (Medvedev \& Menou 2001).}
\end{figure}

\end{document}